\documentclass[twocolumn,showpacs,superscriptaddress,prb]{revtex4-1}
\usepackage{graphicx}
\usepackage{color}
\usepackage{amsmath,amsthm,amssymb}
\usepackage{braket}
\usepackage[colorlinks,citecolor=blue,linkcolor=red]{hyperref}
\begin{document}
 
\title{Competition between electronic correlations and hybridization in CaMn$_{2}$Bi$_{2}$}

\author{Christopher Lane}
\email{laneca@lanl.gov}
\affiliation{Theoretical Division, Los Alamos National Laboratory, Los Alamos, New Mexico 87545, USA}
\affiliation{Center for Integrated Nanotechnologies, Los Alamos National Laboratory, Los Alamos, New Mexico 87545, USA}

\author{M. M. Piva}
\affiliation{Institudo de F\'{i}sica ``Gleb Wataghin'', UNICAMP, 13083-859, Campinas, SP, Brazil}

\author{P. F. S. Rosa}
\affiliation{Division of Materials Physics and Application, Los Alamos National Laboratory, Los Alamos, New Mexico 87545, USA}

\author{Jian-Xin Zhu}
\email{jxzhu@lanl.gov}
\affiliation{Theoretical Division, Los Alamos National Laboratory, Los Alamos, New Mexico 87545, USA}
\affiliation{Center for Integrated Nanotechnologies, Los Alamos National Laboratory, Los Alamos, New Mexico 87545, USA}

\date{\today} 
\begin{abstract}
We study the interplay between electronic correlations and hybridization in the low-energy electronic structure of CaMn$_2$Bi$_2$, a candidate hybridization-gap semiconductor. Utilizing a DFT+$U$ approach we find both the antiferromagnetic N\'eel order and band gap in good agreement with the corresponding experimental values. We further find that, under hydrostatic pressure,  the band gap is mainly governed by magnetic correlations, whereas hybridization has a greater impact on states at higher band energies. This result suggests that CaMn$_2$Bi$_2$ is more closely related to the high-temperature superconducting cuprates and iron pnictides than the heavy fermion class of materials. Finally, we also find the antiferromagnetic CaMn$_2$Bi$_2$ to be topologically trivial for all pressures studied. 
\end{abstract}

\pacs{}

\maketitle 

\section{Introduction}
The electronic structure of fermionic correlated systems is driven by the competition between the tendencies of the electron to spread out as a wave and to localize as a particle, the latter usually accompanied with magnetism. That is, the interplay of the spin and charge degrees of freedom is a central issue.~\cite{nagaosa1999quantum} Layered two-dimensional (2D) materials provide a unique platform for studying this dual nature of the electronic states which produces rich phase diagrams including high temperature superconductivity,~\cite{stewart2011superconductivity,proust2019remarkable} non-trivial topological insulating and semi-metallic phases,~\cite{bansil2016colloquium} and quantum spin liquid states.~\cite{takagi2019concept}

In particular, the iron-based superconductors have been under vigorous experimental and theoretical study since the discovery of unconventional high-temperature superconductivity in La[O$_{1-x}$F$_x$]FeAs in 2008.~\cite{kamihara2008iron} Since then a family of compounds with related layered crystal structures and chemical compositions were discovered including FeSe, LiFeAs, $R$FeAsO ($R$=rare earth), $A$Fe$_2$As$_2$ ($A$=Ca, Sr, Ba, Eu), termed the `11', `111', `1111', and `122' type structures, respectively.~\cite{wen2011materials} The highest  superconducting transition temperature  of 56 K has been found in the 1111-type compound Gd$_{0.8}$Th$_{0.2}$FeAsO.~\cite{wang2008thorium}

To enhance the superconducting transition temperature and search for new broken symmetry phases, Fe was substituted away and replaced by other transition metals such as Cr, Mn, Co, and Ni. These isostructural compounds form new ground states including metallic (Co-based), itinerant antiferromagnetic (Cr-based), superconducting (Ni-based), and semiconducting antiferromagnetic (Mn-based) behavior. The Mn-based pnictides garnered special interested due to their similarity to the phenomenology of the high-temperature cuprate superconductors. In particular, the Mn-based compounds exhibit insulator-metal transitions upon either doping or application of pressure, but superconductivity has yet to be reported. \cite{yanagi2009antiferromagnetic,mcguire2016short,zhang2016structure,simonson2012antiferromagnetic,simonson2011gap,sun2012insulator,hanna2013antiferromagnetic} In general this suggests that the manganese pnictides possibly form a bridge between the pnictide and cuprate material families.

Recent experimental and theoretical studies find CaMn$_2$Bi$_2$ to host many intriguing properties including large anisotropic magnetoresistance\cite{kawaguchi2018nonmonotonic} and a plane-to-chain structural transition\cite{gui2019pressure}, but most curiously it has been suggested that CaMn$_2$Bi$_2$ may be a hybridization-gap semiconductor.~\cite{gibson2015magnetic,zada2019first} In line with this claim, low-temperature electrical transport measurements find a slight increase of the gap under pressure.~\cite{piva2019putative} This type of behavior is akin to Ce$_3$Bi$_4$Pt$_3$ and other heavy fermion compounds.~\cite{hundley1990hybridization,cooley1997high,campbell2019high} Therefore, CaMn$_2$Bi$_2$ could provide a link between the cuprates, pnictides, and heavy fermion systems. 

In this article, we present a first-principle investigation of the electronic and magnetic structure of CaMn$_2$Bi$_2$. We find electronic correlations to dominate the band gap, with the effect of hybridization limited to energies 0.5 eV and higher. In the pristine case, we are able to obtain an accurate ground state by including an effective Harrbard $U$, which significantly improves the agreement with experiments over earlier theoretical studies where GGA-PBE predicts a metal,~\cite{gibson2015magnetic} while hybrid functional dramatically overestimates the gap by an order-of-magnitude.~\cite{piva2019putative} The good agreement also provides an important starting point for the study of pressure effects. Under applied hydrostatic pressure we find an insulator-metal transition near 10 kbar along with reduced manganese magnetic moments as expected for a correlated system. Hybridization is seen to increase with pressure, but the energy scales at which it plays a role prevent it from influencing the band gap. Finally, we also find the antiferromagnetic CaMn$_2$Bi$_2$ to be topologically trivial for all studied pressures . 

{\it Ab initio} calculations were carried out by using the pseudopotential projector-augmented wave method~\cite{Kresse1999} implemented in the Vienna ab initio simulation package (VASP)~\cite{Kresse1996,Kresse1993} with an energy cutoff of $600$ eV for the plane-wave basis set. Exchange-correlation effects were treated using the Perdew-Burke-Ernzerhof (PBE) GGA density functional,~\cite{perdew1996generalized} where a 12 $\times$ 12 $\times$ 8 $\Gamma$-centered k-point mesh was used to sample the Brillouin zone. Spin-orbit coupling effects were included self-consistently. We used the low-temperature $P\bar{3}m1$ (164) crystal structure in accord with the experimental measurements.~\cite{cordier1976neue} For each $U$ and pressure, all atomic sites in the unit cell along with the unit cell dimensions were relaxed simultaneously using a conjugate gradient algorithm to minimize energy with an atomic force tolerance of $0.01$ eV/\AA~and a total energy tolerance of $10^{-6}$ eV. The theoretically obtained structural parameters are in good agreement with the corresponding experimental results.

\section{Magnetic and Electronic Structure}

\begin{figure}[b]
\includegraphics[width=.99\columnwidth]{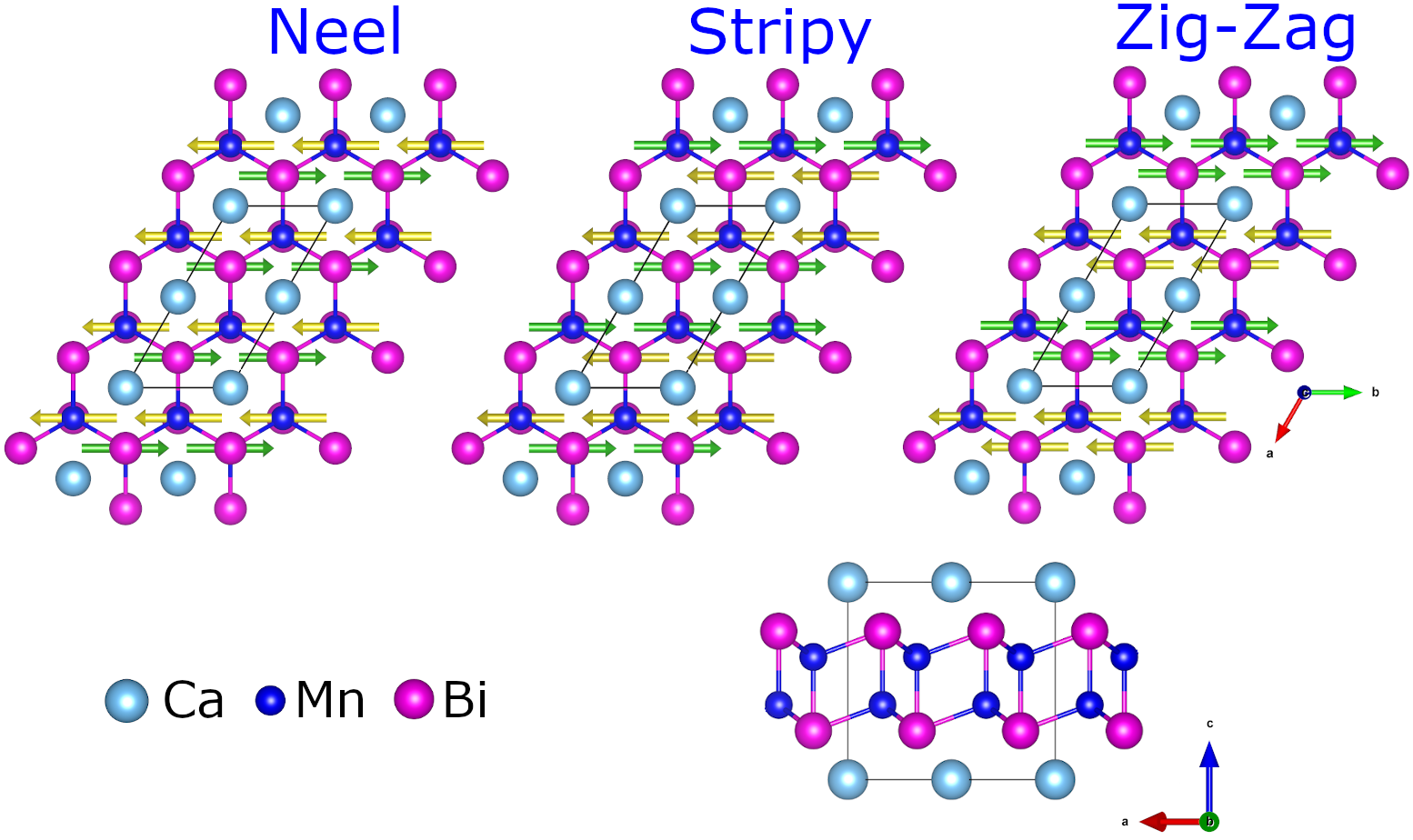}
\caption{(color online) Various antiferromagnetic ground state configurations within the crystal structure of CaMn$_2$Bi$_2$.  Green (gold) arrows represent the positive (negative) manganese magnetic moments. The stacking of the atomic layers is shown on the lower right. The black lines mark the unit cell. } 
\label{fig:MagArrangements}
\end{figure}

Figure~\ref{fig:MagArrangements} shows the three possible antiferromagnetic ground state configurations within the crystal structure of CaMn$_2$Bi$_2$. The magnetic moments (green and gold arrows) are stabilized on the manganese sites within the plane oriented along the $b$-axis in accord with experimental observations.~\cite{gibson2015magnetic} Our first principles total energy calculations find the N\'eel-type order to be the ground state consistent with neutron diffraction,~\cite{gibson2015magnetic} with the other candidate magnetic states lying  at least 138 meV above in energy. The magnitude of the magnetic moments along with the band gap and relative total energy of the various magnetic configurations are given in Table~\ref{table:groundstate}. Experimentally, the N\'eel phase exhibits a magnetic moment of 3.85 $\mu_{B}$ and a band gap between 31 - 62 meV, depending on the report~\cite{gibson2015magnetic,sangeetha2018antiferromagnetism}. Additional recent transport studies find a small activation gap between 2 - 4 meV.\cite{piva2019putative} Our PBE-based calculations yield a magnetic moment close to the experimental value, but with a zero energy gap. \footnote{We have also tested the new state-of-the-art density meta-GGA functional SCAN  and find a slightly enhanced magnetic moments with a concomitant $\sim 200$ meV band gap. Curiously, this enhancement in band gap and magnetic moments is not found in studies of the cuprates,~\cite{Furness2018,Lane2018,zhang2020competing,nokelainen2020emph,pokharel2020ab} iridates,~\cite{lane2020first} and 3d perovskite oxides in general.~\cite{varignon2019origin} }

\begin{table}[ht!]
\caption{\label{table:groundstate}Comparison of various theoretically predicted  properties for the three possible  antiferromagnetic ground states in CaMn$_2$Bi$_2$. }
\begin{ruledtabular}
\begin{tabular}{cccccc}
Order & Magnetic  & Orbital & Total & Gap & Relative Energy\\
& ($\mu_{B}$) & ($\mu_{B}$) & ($\mu_{B}$) &  (meV) & (meV/Mn)\\
\hline\hline
DFT &&&&&\\
N\'eel & 3.923 & 0.103 & 4.026 & 0 & 0\\
Stripy & 3.926 & 0.081 & 4.007 & 131& 138\\
Zig-Zag & 3.981 & 0.084 & 4.065 & 0 & 431\\
DFT+U &&&&&\\
N\'eel    & 4.059 & 0.094 & 4.153 & 32  & 0   \\ 
Stripy  & 4.055 & 0.075 & 4.13 & 194 & 125 \\ 
Zig-Zag & 4.097 & 0.078 & 4.175 & 80  & 373 \\ 
\end{tabular}
\end{ruledtabular}
\end{table}

\begin{figure*}[t]
\includegraphics[width=.99\textwidth]{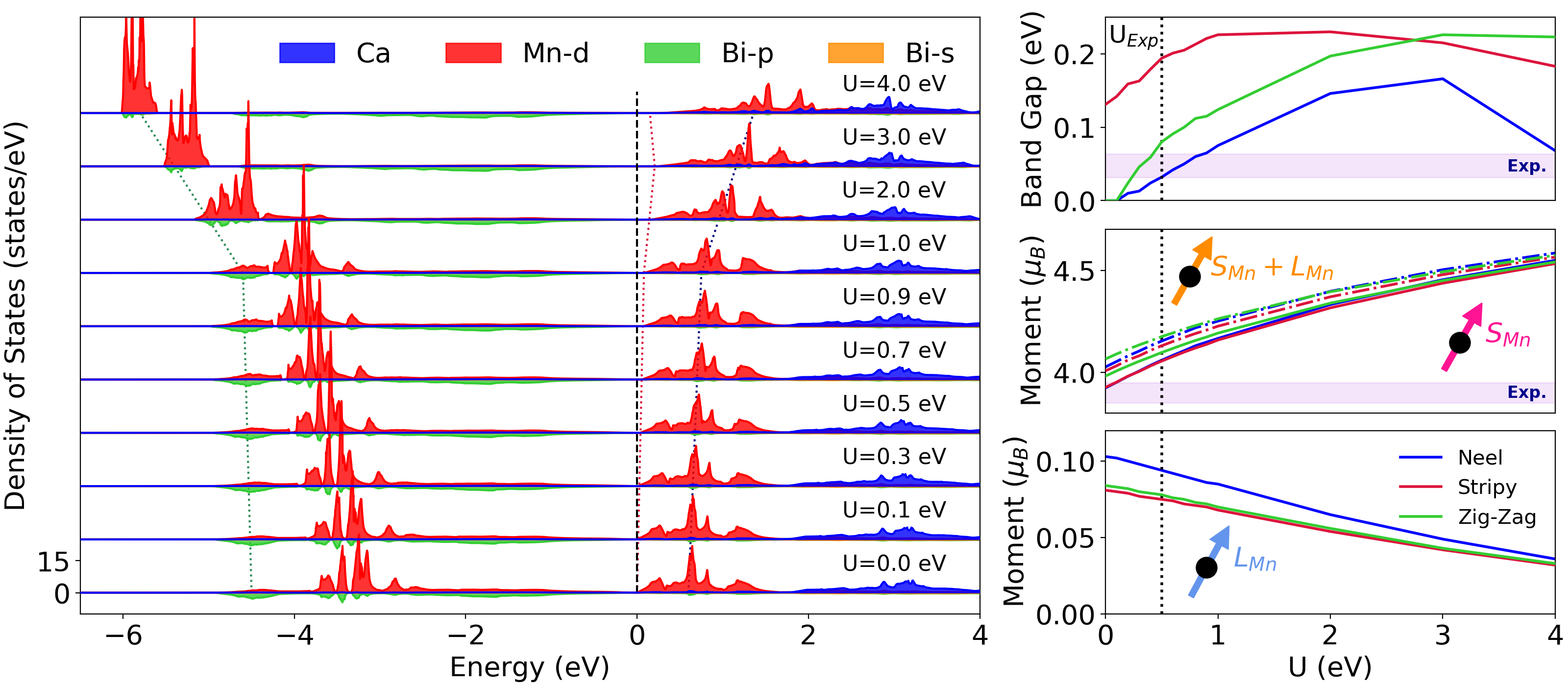}
\caption{(color online) (left panel) Site-resolved partial density of states for the N\'eel AFM phase for various values of $U$. Shading and lines of various colors (see legend) give the contributions from manganese-$d$, bismuth-$s$ and -$p$ orbitals, and Ca atomic weight. The green (blue) dashed lines follow the shift in bismuth (manganese) states with increasing $U$. The red dashed line marks the leading conduction band edge, with the black dashed line marking the Fermi Energy. (right panel) The spin and orbital (solid lines) components of the total (dot-dashed lines) magnetic moment, along with the band gap as a function of the on-site potential $U$.} 
\label{fig:variousU}
\end{figure*}

To remedy the underestimation of the band gap, we introduce an effective $U$ on the Mn-$d$ states within the GGA+$U$ scheme of Dudarev et al.~\cite{dudarev1998electron} To find the $U$ that yields the experimental values, a range of on-site potentials were considered. Figure~\ref{fig:variousU} (right panel) shows the evolution of the band gap, the total (dash-dot lines), and spin and orbital (solid lines) magentic moments as a function of $U$ for the various magnetic configurations, along with the average experimental values overlaid (violet shading). Increasing $U$ from 0.0 eV to 2.0 eV a band gap opens in the N\'eel and Zig-Zag phases, while the gap in the Stripy arrangement increases monotonically. For $U$ greater than 2.0 eV, the gap in the Stripy and N\'eel orders decrease, while that of the Zig-Zag phase flattens. Simultaneously, the strength of the spin (orbital) magnetic moments continually increase (decrease) with increasing $U$. Since the finite orbital moment induced on the manganese atom is mainly via a hybridization with the p-orbitals on the bismuth atoms with strong spin-orbit coupling, the decrease in this moment with $U$ shows clearly a  correlation-induced reduction of hybridization between the atomic species, in favor of electron localization on the Mn site. This localization process concomitantly drives the increase in the size spin component of the moment. A $U_{Exp}$ of 0.5 eV is found to reproduce the experimental gap while only slightly enhancing the magnetic moment, as indicated by the black dotted line. These GGA+$U$ results improve upon those obtained using the HSE06 hybrid functional. \cite{piva2019putative}

Figure~\ref{fig:variousU} (left panel) shows the site-resolved density of states (DOS) for various atomic orbitals including Mn-$d$, Bi-$s$,$p$, and total Ca weight as a function of the effective $U$.  On tuning $U$ from 0 eV to 4 eV the gap in the magnetically split Mn-$d$ states (centered at -3 eV and 1.0 eV) clearly expands, marking a rise in the Mn-$d$ on-site correlations. Below the Fermi energy, the overlap between Bi-$p$ and Mn-$d$ orbitals is diminished, due to the reduction in Mn-$d$ weight, while the bands move away from the Fermi energy.  Above the Fermi level, the center-of-gravity  of the Mn-$d$ states shifts towards higher energies with increasing $U$; however the bandwidth stays relatively the same, which implies no significant change in Bi/Mn hybridization. The Ca dominated band stays centered about 3 eV above the Fermi energy for all $U$, illustrating that $U$ is just influencing the correlation strength of the Mn levels.  Additionally, we find the Bi-$s$ orbital to have marginal weight throughout the energy range discussed due to its highly delocalized nature.

\begin{figure}[b]
\includegraphics[width=.99\columnwidth]{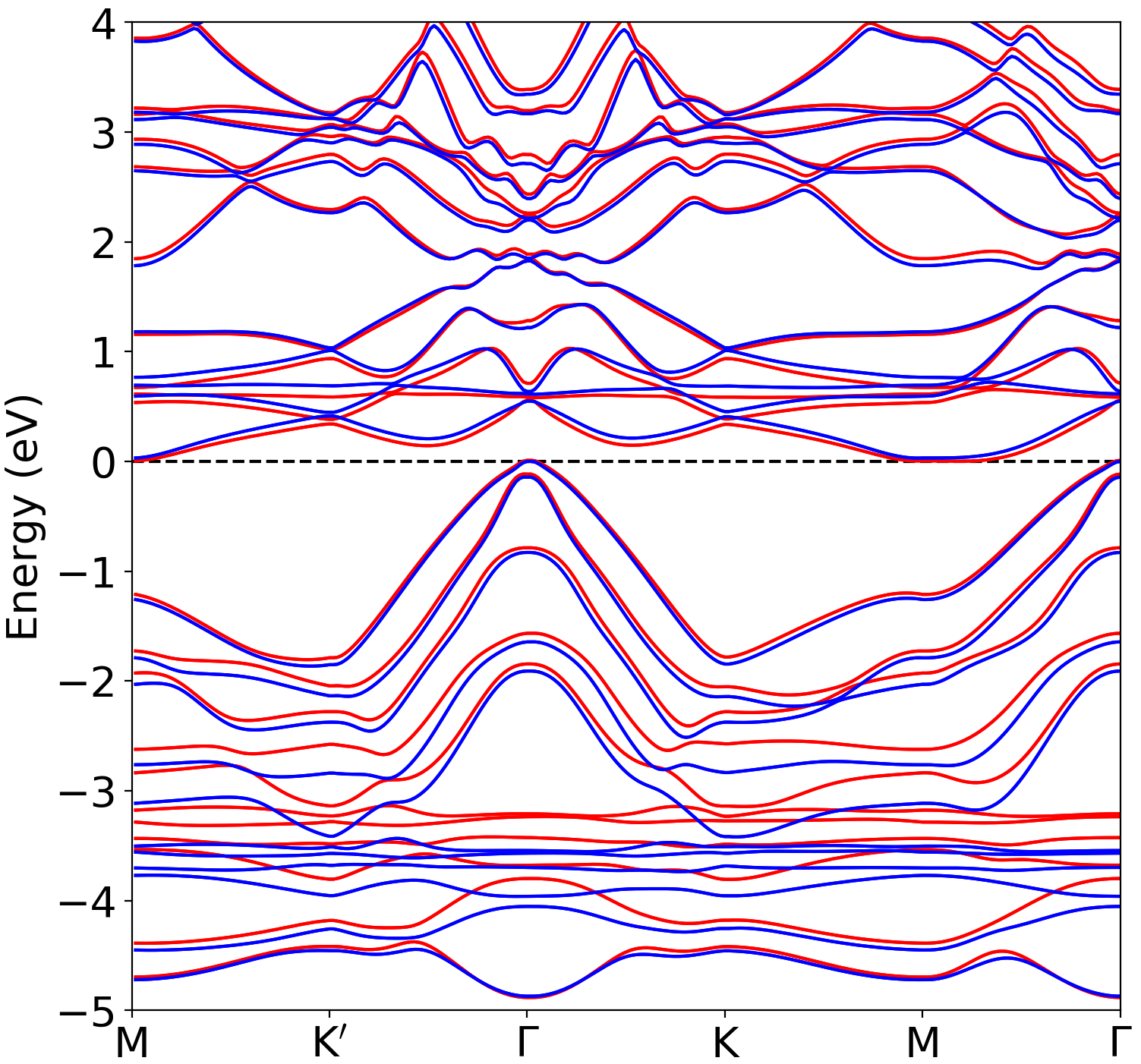}
\caption{(color online) Electronic band dispersion of CaMn$_2$Bi$_2$ in the N\'eel AFM phase with (blue lines) and without (red lines) an effective on-site potential $U_{Exp}$ of $0.5$ eV.} 
\label{fig:Bands}
\end{figure}

Figure~\ref{fig:Bands} shows the electronic band structure of CaMn$_2$Bi$_2$ in the N\'eel AFM phase with (blue) and without (red) the effective Hubbard $U_{Exp}$ of $0.5$. The band gap is indirect with the transition from the conduction band and the valence band characterized by a change in crystal momentum from $\Gamma$ to $M$. Overall, the band structure resembles the same obtained by the bare generalized gradient approximation\cite{piva2019putative,gibson2015magnetic}, except with a finite band gap. In contrast, the conduction bands obtained by the HSE06 hybrid functional (Ref. \onlinecite{piva2019putative}) are more dispersive and display characteristically different transitions and should be noticeable in optical spectroscopy.

\begin{figure*}[t]
\includegraphics[width=.99\textwidth]{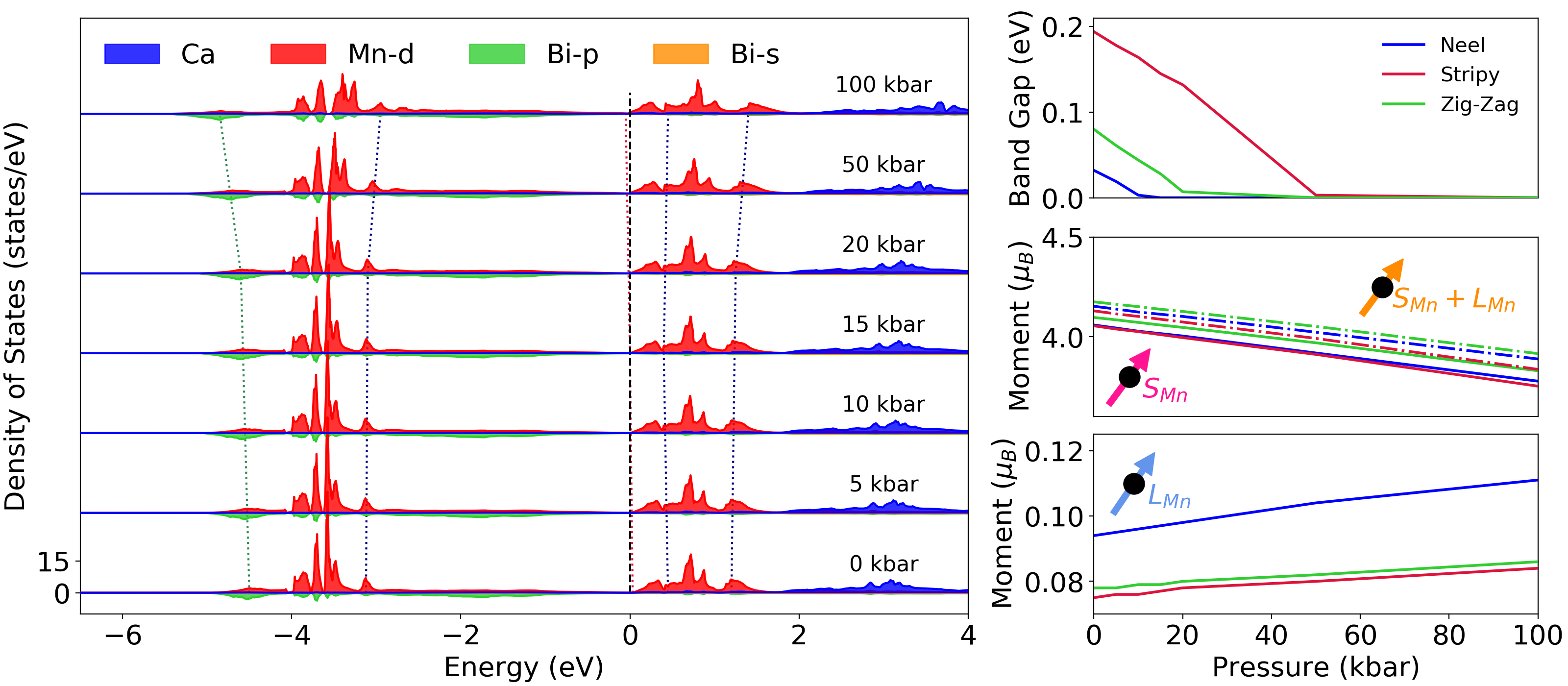}
\caption{(color online) (left panel) Site-resolved partial density of states in the N\'eel type antiferromagnetic phase of CaMn$_2$Bi$_2$ under various values of pressure. Shading and lines of various colors (see legend) give the contributions from manganese-$d$, bismuth-$s$ and -$p$ orbitals, and total Ca atomic weight. The green (blue) dashed lines follow the shift in bismuth (manganese) states with increasing pressure. Red solid line tracks the leading edge of the conduction states as the system passes through a metal-insulator transition.  Black dashed line marks the Fermi Energy. (right panel) The spin and orbital (solid lines) components of the total (dot-dashed lines) magnetic moment, along with the band gap as a function of pressure.} 
\label{fig:Pressure}
\end{figure*}

We further note that upon introducing $U$ the relative energy between the different AFM ground state configurations has changed. Now we find the N\'eel and Stripy types to be separated by $\sim 100$ meV, while the energy difference to the Zig-Zag ordered state is $\sim 400$ meV, making the Stripy and Zig-Zag phases irrelevant in the ground state. For more details see Appendix~\ref{sec:totalenergy}.

\section{Effect of Pressure}
Applied external pressure provides a direct means to gauge the relative ratio between hybridization and correlation strengths in a material. If the band gap is driven by correlations, pressure squeezes the lattice sites of the crystal closer together forcing the wave functions of neighboring atomic sites to overlap. Electrons then tend to become more delocalized in the material, yielding a metal. In contrast, if the band gap is governed by hybridization, pressure further separates bonding and anti-bonding states, thus increasing the band gap. When both hybridization and correlations are present, the gap can undergo non-monotonic behavior under pressure as a result of their competition.

Figure \ref{fig:Pressure} (left panel) shows the site-resolved partial density of states in the N\'eel-type antiferromagnetic phase of CaMn$_2$Bi$_2$ under hydrostatic pressures from 0 to 100 kbar.  At zero pressure the AFM order opens a gap in the Mn-$d$ states of $32$ meV, consistent with the range of reported experimental values, along with the bismuth states centered at $1.5$ and $-4.5$ eV, respectively. As pressure is applied, the energy separation between the bismuth states increase, as indicated by the green and outer blue dashed lines, due to the enhanced hybridization. The motion of the manganese states is not as clear-cut. The manganese dominated levels below $E_{F}$ move to towards higher energies, reducing the band gap. There is also a concomitant broadening of the states and a strengthening of Mn-$d$ and Bi-$p$ overlap. 

In contrast, the manganese levels above $E_{F}$ exhibit anomalous non-monotonic behavior. For pressures less than 20 kbar, the peaks in the DOS stay relatively constant in energy, only showing a slight shift to the left (inner blue dashed line).  For pressures greater than 20 kbar, the motion of peaks in the DOS appears to be proportional to their proximity to the Fermi level. For peaks at 0.5 eV or greater, the DOS move to right, following the bismuth $s$-orbital  states, whereas states closer to the Fermi energy move to the left. The leading edge of the conduction band follows a linear trajectory (red dashed line), crossing the Fermi level at  10 kbar, making CaMn$_2$Bi$_2$ a metal. 

This behavior can be rationalized as follows. For small pressures hybridization dominates, pushing states to higher energies, but for larger pressures the collapse of the magnetic correlations is quicker than the change in the hybridization, pulling the states to the left. Therefore, the hybridization plays a dual role by trying to enlarge the gap near the Fermi level, while simultaneously killing magnetic correlations; annihilating the gap. Here we can conclude that the gap appears to be mostly governed by correlations rather than hybridization.  However, due to the anomalous behavior of the Mn-$d$ levels, a slight perturbation or tuning of the localization of the Bi-$s$ level might result in greater hybridization influence on the gap, boosting the band gap before closing at higher pressures.

Figure~\ref{fig:Pressure} (right panel) shows the total magnetic moment along with its spin and orbital components and the band gap as a function of pressure for the various magnetic orders. The total magnetic moment is dominated by spin magnetization both displaying a monotonic linear decrease with pressure. On the other hand, the orbital contribution increases in a power law fashion, with a slight  plateau upon the gap closing.  Here, the the competition between hybridization and correlation with pressure can be readily observed. The spin magnetization provides a direct indicator of the strength of correlations, while the orbital component tracks the overlap of Bi and Mn atomic wave functions, which in turn induce an effective spin-orbit coupling on the Mn sites. Additionally, we also find the Mn moments to slightly ($\approx 1^{\circ}$) tilt out-of-plane with applied pressure.

In the original analysis performed by Gibson {\it et al.}\cite{gibson2015magnetic} it was claimed that the gap behaves as a hybridization gap. This was justified by tracking the changes in position of the Bi-$p$ and $s$ levels with expanding and contracting the unit cell volume by 1\% without relaxation. However, after cross checking our results results within VASP and Wien2k\cite{blaha2001wien2k} we find the electronic structure presented by Gibson {\it et al.} to be inconsistent. We believe the bands presented in Ref.~\onlinecite{gibson2015magnetic} resulted from an incomplete self-consistent process where the spin-orbit coupling subroutines were possibly neglected. Additionally, the pressure study by some of us~\cite{piva2019putative} reports an increase in activation energy of 20 K to 40 K (2 - 4 meV) with pressure using electrical transport measurements. The gap observed in Ref.~\onlinecite{piva2019putative} is one order of magnitude smaller than other reported values,~\cite{gibson2015magnetic} implying the possible existence of impurity states within the gap. The presence of impurities would make the sample sensitive to changes in external pressure and could produce anomalous transport results. Therefore, to accurately compare our first-principles results with the experimental measurements a rigorous modeling of the transport process is required.\footnote{Additionally, Ref.~\onlinecite{piva2019putative} reported an increase in $T_N$ with pressure. However, since our density functional theory results only produce the magnetic moment at zero temperature without any measure of the fluctuations, we cannot give any direct insight into the change in $T_N$ with pressure. Nevertheless, the reduction and reversal in the relative energy between the N\'eel and Zig-Zag AFM phases suggest the presence of strong fluctuations which may drive the slight increase in $T_N$. }

\section{Topological Character}
Originally, CaMn$_2$Bi$_2$ was thought to be a possible magnetic 3D Dirac semimetal, where the Mn-$d$ states were assumed to behave as core electrons. \cite{zhang2019catalogue} This then allows for a clean band inversion of Bi-$s$ and Bi-$p$ levels. However,  Gibson {\it et al.}~\cite{gibson2015magnetic} found that the Mn-$d$ orbitals play a significant role at the Fermi level, hybridized with the manifold of bismuth states. This ultimately disrupts the Bi-$s$ and Bi-$p$ level, avoiding a topological non-trivial ground state. 

To confirm the topological nature of CaMn$_2$Bi$_2$ we used the vasp2trace code~\cite{vergniory2019complete} in conjunction with the Check Topological Material module~\cite{bradlyn2017topological,vergniory2017graph,elcoro2017double} provided on the Bilbao Crystallographic Server.~\cite{aroyo2011crystallography,aroyo2006bilbaoI,aroyo2006bilbaoII} CaMn$_2$Bi$_2$ is indeed found to be topological trivial for all pressures up to 100 kbar, due to the fact that the Mn-$d$ levels dominate the low energy electronic structure and preventing the Bi-$s$ and $p$ states to overlap and invert. However, if the on-site energy of the manganese bands were to be tuned away from the Fermi level or the Bi-$s$ state brought closer to the Fermi level, the bismuth $s$ and $p$ level could be inverted,  making the ground state topologically non-trivial.


\section{Concluding Remarks}\label{sec:conclusion}
By examining the ground state electronic structure of CaMn$_2$Bi$_2$ as a function of pressure, we find the low energy electronic structure to follow the characteristic behavior of a correlation gap. This behavior implies CaMn$_2$Bi$_2$ is more closely related to the cuprate high-temperature superconductors than the Ce$_3$Bi$_4$Pt$_3$ heavy fermion compounds. To fully elucidate its connection to the cuprate compounds and to what extent they are similar, i.e. exhibiting charge/spin density waves and superconductivity, further doping dependent studies are needed to uncover its full phase diagram. Moreover, the addition of holes should produce an interesting interplay between itinerant antiferromagnetic carriers and those on the strongly spin-orbit coupled Bi-$p$ orbitals, creating a ripe environment for new exotic phases of matter.


\begin{acknowledgments}
This work was supported by the U.S. DOE NNSA under Contract No. 89233218CNA000001 and by the Center for Integrated Nanotechnologies, a DOE BES user facility, in partnership with the LANL Institutional Computing Program for computational resources. Additional support was provided by DOE BES Core  Programs (LANL Codes: E3B5 and E1FR). M. M. Piva acknowledges the S\~ao Paulo Research Foundation (FAPESP) grants
2015/15665-3, 2017/25269-3, 2017/10581-1. 
\end{acknowledgments}

\appendix

\section{Relative Energy between Various Magnetic Phases}\label{sec:totalenergy}
\begin{figure}[t]
\includegraphics[width=.99\columnwidth]{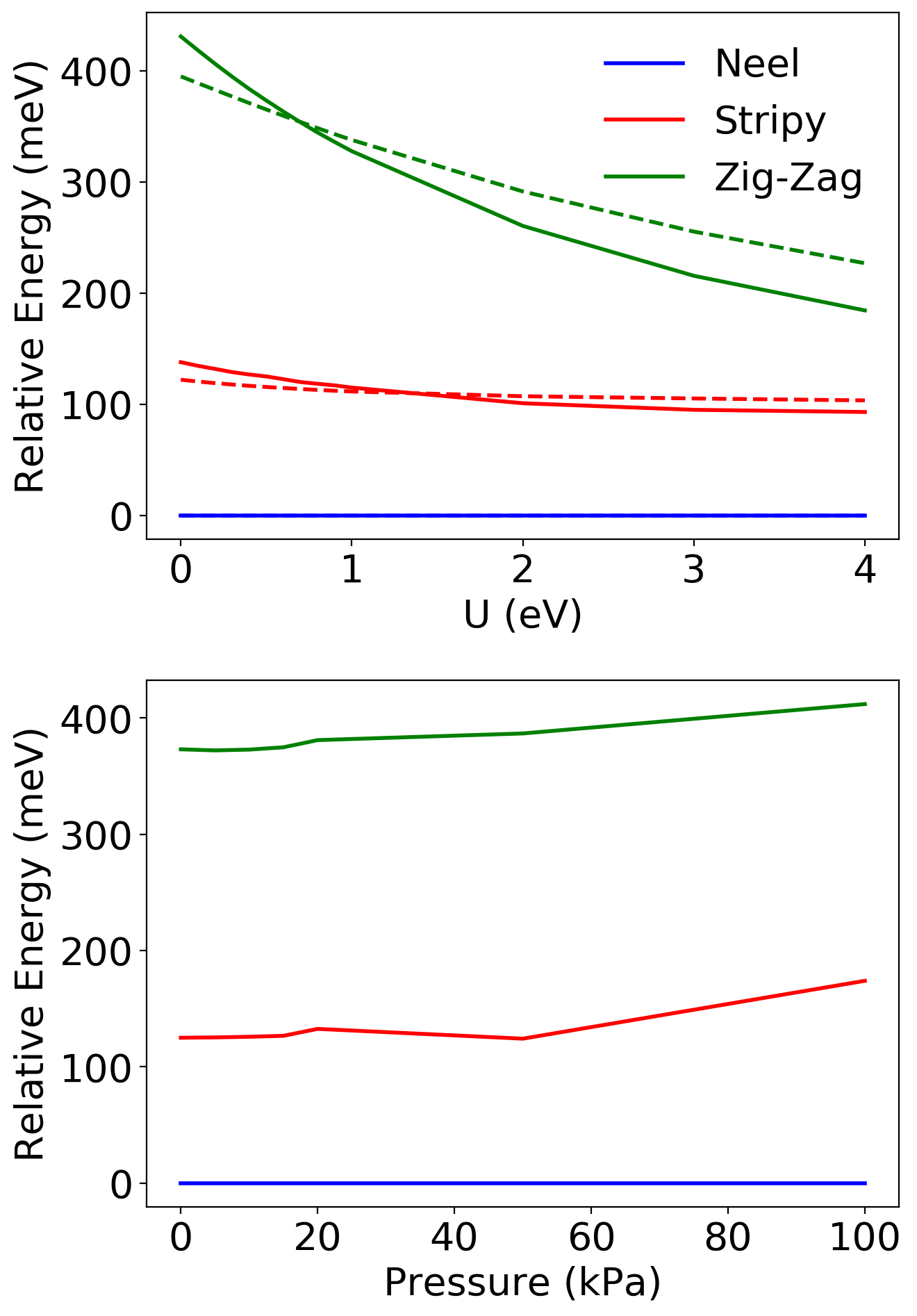}
\caption{(color online) Relative energy of N\'eel, Stripy, and Zig-Zag magnetic phases as a function of Hubbard $U$ (top panel) and pressure (bottom panel). The solid lines (dashed lines) are for the fully relaxed (experimental) crystal structure. } 
\label{fig:Relenergy}
\end{figure}

Figure~\ref{fig:Relenergy} shows the relative energy of various magnetic phases as a function of Hubbard $U$ and pressure. For all pressures (U) studied the N\'eel AFM order is the ground state, with the  Stripy and Zig-Zag configurations lying $\sim 100$ meV and $\sim 300$ meV above, respectively.

\bibliography{CaMn2Bi2_Refs}

\end{document}